# Spin-Coated Erbium-Doped Silica Sol-Gel Films on Silicon

S. Abedrabbo[1,2,3], B. Lahlouh[3], S. Shet[4], A. T. Fiory[1], and N. M. Ravindra[1]

[1]Department of Physics, New Jersey Institute of Technology; Newark, NJ 07901, USA
[2]Department of Physics and Engineering Physics, Stevens Institute of Technology; Hoboken NJ 07030 USA
[3]Department of Physics, University of Jordan; Amman 11942, Jordan
[4]National Renewable Energy Laboratory; Golden CO 80401, USA

## Abstract

This work reports optical functionality contained in, as well as and produced by, thin film coatings. A sol-gel process, formulated with precursor active ingredients of erbium oxide and tetraethylorthosilicate (TEOS), was used for spin-coating thin (~130 nm) erbium-doped (~6 at. %) silica films on single-crystal silicon. Annealed films produce infrared emission in the 1.5-µm band from erbium ions in the film, as well as greatly enhancing (~100X) band-gap emission from the underlying silicon. The distinctly different mechanisms for the two modes of optical activities are interpreted in terms of optical emission theory and modeling; prospects for opto-electronic applications are discussed.

## 1. Introduction

Spin-coating of sol-gel films is widely applied as a cost-effective process for depositing nominally pure and impurity-doped thin silica films on substrates (see e.g. [1]). Of interest are optically active coatings suitable for silicon-integrated optoelectronics. The present method entails depositing erbium-doped silica films on silicon wafer material that are densified by thermal annealing. With suitable selection of thermal annealing temperature ($T_a$), photoluminescence measurements show strong emission in the 1.5-µm wavelength band from $Er^{+3}$ centers in silica (Stark-split intra-4f $4I_{13/2}$ - $4I_{15/2}$ transitions of $Er^{+3}$ [2]) for $T_a \approx 850$ °C [3]; additionally, near band-gap emission at 1.16 µm from the silicon becomes greatly enhanced for $T_a \approx 700$ °C [4]. Section 2 describes experimental procedures and reports analysis of photoluminescence from $Er^{+3}$ in the sol-gel material. Section 3 presents analysis of photoluminescence from the silicon. General conclusions are presented in Section 4.

## 2. Optical Emission from Erbium

Optical telecommunication networks use rare-earth doped optical materials for a variety of applications, e.g. fiber and waveguide amplifiers, waveguides by index of refraction modification, infrared light sources, integrated optical devices, displays and lasers [2,5-8], and silica doped with erbium is a widely used optical material. Although $SiO_2$:Er can be prepared in film form by a variety of methods, spin-coating of sol-gels acquired special interest because of low deposition temperature and low overall cost [6,7,9-13]. For this work, a sol-gel technique was developed using $Er_2O_3$ as the Er dopant source and introduced at high concentration (6 at.%) in the $SiO_2$:Er film. The process yields optically active erbium-doped silica films at moderate annealing temperatures ($T_a$ ~850 °C). An example of a suitable optoelectronic application is in fabrication of planar erbium-doped waveguide amplifiers (EDWAs).

Erbium Sol-Gel Processing

The erbium-doped sol was produced by hydrolysis of TEOS (tetraethylorthosilicate, $Si(OC_2H_5)_4$) in a solution containing erbium oxide and then annealing the gel. The relatively high Er concentration of 6 at.% is expected to form stable Er-O:Si-O structures, owing to the strong Er-O bond (6.3 eV [14]) and the role of Er as a network element in glasses [15,16]. Special considerations of sol-gel processes concern quenching of the 1.5-µm emission by resonant energy transfer to vibrations of residual OH [1,17] and dependence of external emission on Er concentration [10,11]. Much residual OH can be removed with high-temperature annealing (e.g. $T_a$ ~ 900 – 1100 °C [8,18]); however, $T_a$ < 1000 °C is preferable for integrated silicon-based optoelectronics and glass reflow processing [19]. In the process developed for the present work, $T_a$ ~ 850 °C turns out to be sufficient for removing the effects of residual OH as well as reducing surface pores area to maintain low level of OH, thereby avoiding high-temperature processing.

Experimental Procedure

The starting solution for the sol-gel process contained 0.5 g $Er_2O_3$ powder mixed into a solution of 4 ml ethanol, 4 ml acetic acid, and 1.6 ml deionized water that was stirred at 45 °C for 3 hours. The sol was then prepared by the addition of 2 ml TEOS and stirring for 10 min at 80 °C. Hydrolysis with high water/TEOS molar ratio R (here, R = 10) with acid catalysis produces branched Si-O polymerization in the sol and leads to dense films [20]; the trade-off is an increased hydroxyl content to be removed by thermal annealing [13]. Following this step, the solution was passed through a syringe filter with 0.45-µm pore size and spun coated on 2.5-cm pieces of cleaned Si (100) wafer substrates rotating at 1200 rpm for 30 seconds. The resulting gel films were then oven dried in air at 120 °C for 30 minutes. Post-deposition thermal treatment was studied by vacuum annealing (2 Pa) at temperatures ($T_a$) spanning the range 500 – 950 °C for a duration of one hour ($t_{anneal}$). Prior work has shown that vacuum annealing is more effective in removing OH contaminants than annealing in air, as determined from increased $Er^{+3}$ emission lifetime [13].

Film thickness and index of refraction were determined using a thin-film spectrophotometer. A model Fluorolog-3 spectrofluorometer (Horiba Jobin Yvon) was used to obtain room temperature photoluminescence (PL) emission from the $Er^{+3}$ centers in the 1535 nm band. A Xe lamp was used for excitation with a double excitation monochromator to set a fixed excitation wavelength in the range of 515 to 530 nm. Spectral signal intensities were recorded in the range 1400 to 1600 nm using a LN cooled Hamamatsu InGaAs photodiode detector, preceded by a single emission monochromator, using 0.2 s integration time at each wavelength. Emission from $Er^{+3}$ in the 1535 nm band was maximized by tuning the excitation monochrometer to 521 – 523 nm, as obtained from a three-dimensional matrix of photoluminescence-excitation spectra. Photoluminescence spectra were normalized to the power output of the excitation source, monitored by a separate photodiode.

Annealing is generally required to densify sol-gel films by removing water, organic compounds and hydroxyl residues [8,18]; most volatiles are driven off at $T_a$ ~ 500 °C [10]. Films may densify by ~25%, observed in this work as shrinkage in thickness from ~170 nm (air-dried) to

~130 nm ($T_a$ ~ 700 °C ). The $T_a$ for producing the strongest PL depends on process method as well as Er concentration (e.g. $T_a$ ~ 900 °C are often used for sol-gel films [6,7,21]). Decreased emission at high annealing temperatures can arise from a number causes: Segregation by precipitation of Er or phase separation by clustering, and quenching by Er-Er interactions at high optically active $Er^{+3}$ concentrations, as well as residual OH contamination [22].

Results: Erbium Emission

Photoluminescence spectra for samples annealed at various $T_a$ are presented in Figure 1 (normalized logarithmic intensity scale). Peaks near 1160 nm originate from the silicon substrate (see section 3), while emission peaks near 1535 nm arise from $Er^{+3}$ in the sol-gel film. Erbium emission from the as-deposited air-dried sample is very weak and hardly noticed when compared to annealed samples. This is to be expected, since low temperature baking leaves residues from the sol and an abundance of water and hydroxyls in a low-density porous near-glassy network. As such it is expected that non-radiative recombination strongly competes with the radiative ones, leading to the low PL signal observed.

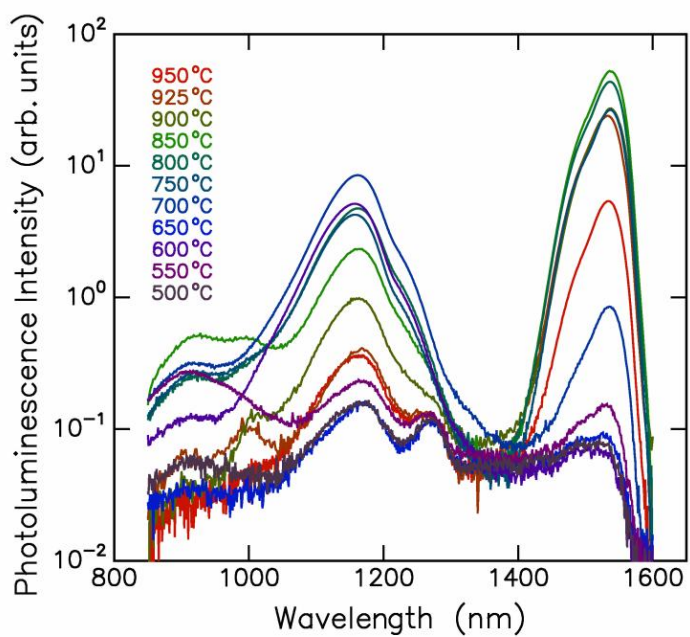

Figure 1. Photoluminescence intensity spectra from silicon coated with erbium-doped sol-gel silica films (anneal temperatures $T_a$ in legend).

The PL signal from the $Er^{+3}$ appreciably improves as a function of increasing $T_a$, particularly for $T_a$ > 700 °C, until it reaches a maximum at $T_a \approx 850$ °C. Samples annealed at higher $T_a$ exhibit a decreasing trend in their PL (confirmed out to 1050 °C for additionally prepared samples annealed in either air or vacuum). Maximum PL intensity occurs at emission wavelengths in the vicinity of 1533 to 1537 nm and spectral full width at half maximum is in the range 51 to 58 nm, both varying somewhat with $T_a$. To illustrate the trend, peak PL intensities (normalized arbitrary units; integrated spectra are similar) are plotted against $T_a$ in Figure 2 as filled circles (uncertainties for anneals above 600 °C are approximately 12% full scale). The evident non-monotonic dependence on $T_a$ is indicative of competing thermal reactions that activate optical emission from the intra 4f $Er^{+3}$ band as $T_a$ is increased towards 850 °C, and de-activate it at the higher $T_a$.

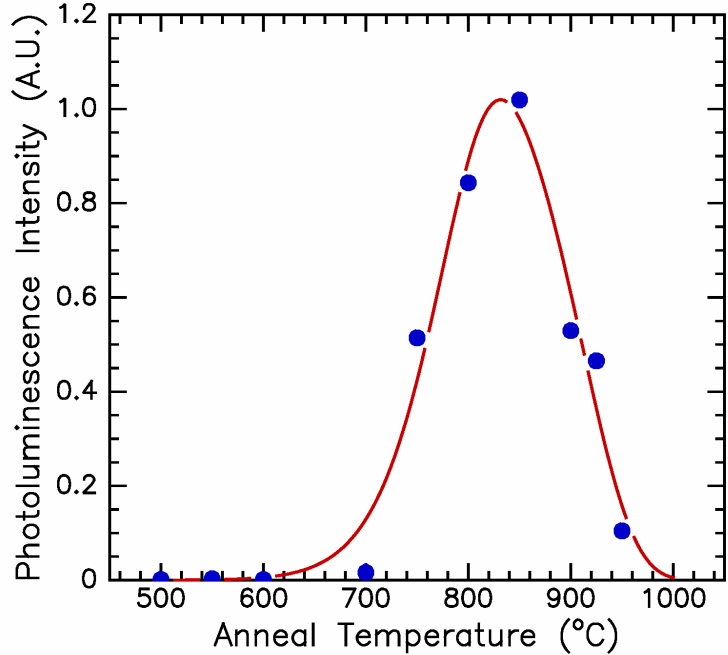

Figure 2. Maximum $Er^{+3}$ photoluminescence intensities (~1535 nm) from silicon coated with erbium-doped sol-gel silica films *vs.* anneal temperature (data: points; theory: curve).

In [3] a model involving two activation energies was introduced to provide a quantitative analysis of the dependence of the photoluminescence from the Er on $T_a$; An energy $E_a$ is defined as the activation energy of the favorable process (forming optically active $Er^{+3}$) and an energy $E_d$ is defined as the activation energy of the unfavorable process (quenching optically active $Er^{+3}$). Fitting the model to the data yields the results $E_a = 2.3 \pm 0.6$ eV and $E_d = 2.6 \pm 0.7$ eV, and the theoretical function shown in Figure 2; error bars in the activation energies arise from parameter correlation and uncertainty in the data. The magnitude of the activation energies is consistent with the idea that OH removal is the mechanism by which the otherwise optically active $Er^{+3}$ yields the PL signal. Interpretations of these results are discussed below.

Influence of Hydroxyl on $Er^{+3}$ Emission

Presence of OH is extremely effective in quenching excited $Er^{+3}$ ions [17]. At high OH concentrations direct resonant energy transfer from the excited ion to OH serves as an extremely effective trap. At low OH concentrations fast transfer of energy from ion to ion via the Förster mechanism (cooperative energy transfer or CET) may allow diffusion of the excitation to an OH impurity, where it becomes dissipated by multi-phonon assisted decay to the oxide host [23]. The latter process is a type of concentration quenching, since the rare-earth concentration is the dominant factor in the CET coefficient [22]. In the present case both mechanisms are expected to apply, since the Er concentration is quite high.

Weak PL emission for the low-temperature anneals ($T_a \ll 850$ °C) indicates quenching of excited $Er^{+3}$ ions by direct Er→OH interactions. Increase in PL intensity in the region $T_a \leq 850$ °C is therefore consistent with removing OH from proximity to $Er^{+3}$ ions. Residual OH is mostly surface bonded within pores in the form of silanol groups, as shown by infrared absorption near 3670 $cm^{-1}$ observed in bulk silica gels (~ 900 °C anneals) [24,25]. An identified mechanism for OH removal is surface desorption of $H_2O$, which is released upon reaction of surface silanol to form siloxane bridge structures, corresponding to an enthalpy $\Delta H = 2.16$ eV [26]. (Bulk studies indicate OH diffuses very rapidly at 850 °C [27].) Since $E_a \approx \Delta H$ within error, one may associate creation of optically active $Er^{+3}$ with the process of OH removal by desorption.

Since the observed diminution of PL intensity at higher annealing temperatures corresponds to a thermal activation energy $E_d$ that is also on the order of that for OH processes, it appears that OH removal is involved in the PL deactivation as well. Reduced concentrations of residual OH obtained for high temperature annealing may allow the diffusive Er→Er→OH processes to become effective for quenching PL emission. Moreover, high Er concentration in itself (e.g. upon virtually eliminating OH and densifying as Er-Er distances decrease) can lead to concentration quenching by cross relaxation or cooperative upconversion processes (CUP) involving Er-Er dipolar interactions, a form of CET, that typically converts one out of two units of excitation energy into heat; this CUP mechanism has been reported for optical fibers with high Er concentrations [28]. Both of these mechanisms can account for the decrease in PL emission for annealing in the region 850 °C $\ll T_a \leq$ 1050 °C.

Diffusivity of Er at low concentrations in various deposited silica films was recently measured in the temperature range 1000 – 1100 °C [29], from which an activation energy $E_{Er} = 5.3$ eV was derived, due mainly to strong Er-O bonds, which are also responsible for stability of Er-glass structures; extrapolation to 950 °C yields diffusivity $D_{Er} \approx 6 \times 10^{-17}$ $cm^2 s^{-1}$ and diffusion length $(4D_{Er} t_{anneal})^{1/2}$ ~ 10 nm. Erbium diffusion therefore appears to be sufficient for Er clustering at 950 °C, owing to the high Er concentration in the sol-gel films (mean Er-Er distance ~ 1 nm); Er clustering generally kills the optical activity of the involved Er ions. Although Er segregation may participate along the other mechanisms leading to overall PL quenching for $T_a$ post 850°C, Er is considered a primary network component at comparable concentrations [15] favoring stable Si-O-Er glassy network.

As in fiber amplifiers doped with Er at high concentration, one expects that Er-Er interaction to saturate the PL but not to shrink it. However, as sol-gel films are usually not completely densified and Er diffusion is negligible for anneals near 850 °C, CET is minimized for $T_a$ ~ 850 °C while OH out-diffusion is very welcomed. Higher annealing temperature permits the Er-Er interaction to occur at much higher rate due to faster densification process where PL deterioration can be spurred by anyone or more of the briefed processes.

These results are quite different from the concentration quenching observed at low Er concentrations. Existence of a range in annealing temperatures where concentration quenching may appear suppressed is a particular advantage obtained by using a high Er concentration. One does not expect thermal annealing behavior to materially depend on Er concentration, since the optimal anneal temperature depends mainly the OH concentration that dominates the process. The optimum $T_a$ at 6 at. % Er is somewhat lower than previously reported for sol-gel films with less than 1 at. % Er [6,7,21]; this is similar to the behavior of Er-doped silicon-rich oxides, where optimal $T_a$ is lowered by about 100 °C at Er concentrations of 3 - 6 at. % [30].

### 3. Enhanced Optical Emission from Silicon

The PL spectra of Figure 1 shows emission at 1.067-eV (1162-nm), which is just below the Si band gap (energy units used herein by convention). Analysis presented below shows that this room temperature emission from silicon is enhanced by the Er-doped sol-gel coatings and is correlated with inhomogeneous film stresses. Obtaining high efficiency in light emission from silicon at room temperature has traditionally entailed selecting structures and materials to circumvent inherent disadvantages of the indirect Si band gap (e.g. typical $10^{-4}$ quantum efficiency at 300 K [31], see also [32]) [33]. Thus this sol-gel process method has inherent advantages as a simplified process.

Results: Enhanced Silicon Emission

Photoluminescence spectra are shown in Figure 3 for two samples that were annealed at $T_a$ of 700 °C and 850 °C (solid and broken curves, respectively) and exhibit major peaks at 0.807 and 1.067 eV. The 1.067-eV peak is strongest for annealing at 700 °C, where it is enhanced by a factor of 50 when compared to unannealed, air-dried films; as verified below, emission at 1.067 eV is associated with the Si substrate.

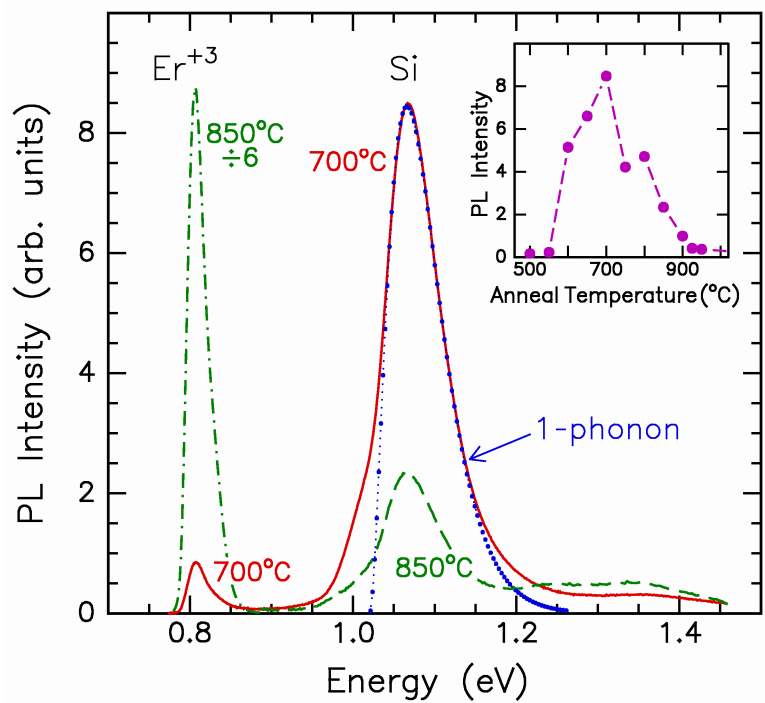

Figure 3. Photoluminescence intensity spectra for two silicon coated with erbium-doped sol-gel silica films, annealed at 700 °C (Si: solid curve) and 850 °C ($Er^{+3}$: dashed; chain-dashed, scaled by 6). Dotted curve denote 1-phonon theory. Insert: peak PL from Si (~1.067 eV) *vs.* anneal temperature.

The 0.807-eV peak is strongest for annealing at 850 °C (left dot-dash curve reduced by factor of 6), and is associated with emission from $Er^{+3}$ in the silica film, as discussed in section 2. Weaker emissions in the region 1.25 – 1.35 eV are attributed to $^4I_{11/2} \rightarrow {}^4I_{15/2}$ transitions in $Er^{+3}$. Annealing at $T_a$ = 850 °C corresponds to maximum 0.807-eV $Er^{+3}$ emission. On comparing the two emission signals, the peak emission from the Si ($T_a$ = 700 °C) is about 16% of the peak emission from $Er^{+3}$ ($T_a$ =850 °C); from areas under the respective spectra, the integrated signal from the Si is 50 % of that from $Er^{+3}$.

The dotted curve in Figure 3 is a theoretical one-phonon (i.e. dominant phonon) model for emission of photons at energy E by phonon-assisted recombination of free electrons and holes in silicon, represented by the expression, $PL(E) \propto (E-E_0)^2 \exp[-(E-E_0)/k_BT$ (uncorrelated pair model, see e.g. [34]); it is calculated with threshold energy $E_0$ = 1.020 eV and thermal energy $k_BT$ = 0.023 eV. Owing to this function, the PL spectra generally exhibit exponential tails for $E>E_0$. Since the dominant phonons are the momentum-conserving transverse optical phonons of energy $E_{ph}$ = 0.0578 eV, the effective band gap, neglecting possible corrections for exciton binding or trapping, is $E_G = E_0 + E_{ph}$ = 1.078 eV, which is about 42 meV below the intrinsic Si band gap (1.12 eV). This is an approximate examination of the data as the effect of weaker phonon components, e.g. features appearing below 1.03 eV, were not included, the reasonable overlap for much of the PL signal peaking at 1.067 eV indicates that this emission originates indeed from the silicon substrate and not the film. X-ray diffraction data show no evidence of polycrystalline Si (e.g. absence of Si nano crystals within the deposited film [33]).

Annealing behavior for emission from $Er^{+3}$ in the sol-gel film and from the Si substrate obey distinctly different dependences on $T_a$ (compare Figure 2 and Figure 3, inset). This shows that emission from the Si does not depend on that from the $Er^{+3}$ ions in the silica film, from which we conclude that emissions at these two wavelengths arise from independent mechanisms.

Influence of Sol-Gel Coating on Si Emission

The observed annealing behavior points to enhancement of emission from the silicon arising from a reversible (owing to non-monotonic dependence on $T_a$) interaction with the sol-gel coating. Stresses in sol-gel films, caused by shrinkage and porosity, are known to be inhomogeneous as indicated by IR absorption signatures (frequency shifted $TO_3$ modes at 1060 – 1080 $cm^{-1}$ [35]), and tend to be greatest for $T_a$ ~ 700 °C [36], which coincidentally is where the strongest Si PL from our samples is observed. Given that Si PL is comparatively subdued for $T_a$ < 600 °C as well as for $T_a$ > 900 °C, one may conclude that enhanced PL at $T_a$ ~ 700 °C is to be associated with non-uniformities in film stresses.

Inhomogeneous film stresses, acting in concert with native interfacial roughness, is therefore expected to create local strain-induced band bending in the silicon and increase the effective cross section for radiative free-carrier recombination. Applying this interpretation, enhanced emission is expected to be produced in only a thin layer at the Si surface (on the order of the film thickness, ~0.1 μm, or less); since the excitation radiation penetrates ~1 μm, quantum efficiency (QE) could actually be increased a thousand-fold (i.e. on the order of ten times larger than the observed enhancement), suggesting QE >1% (assuming QE ≤ 0.01% as reported for CZ silicon [31,32]).

## 4. Conclusions

A process for producing optically active $SiO_2$:Er thin films on Si substrates using low cost sol-gel techniques that utilize $Er_2O_3$ has been presented and analyzed. Photoluminescence is enhanced strongly as a function of annealing temperature, reaching optimum for annealing temperatures in a 50-°C range near 850 °C. External emission from Er-O structures is shielded by OH for annealing temperatures $T_a < 850$ °C and by the combination of Er concentration quenching and Er-Er-OH energy diffusion for $T_a > 850$ °C. The results indicate that minima in quenching by Er-Er cooperative energy transfer and by Er-OH interactions can be associated with $T_a \approx 850$ °C.

In addition, near-band gap (1.067 eV) photoluminescence at room temperature has been observed in CZ Si wafer material with the erbium-doped sol-gel silica coating. Emission from the Si, strongest for $T_a \approx 700$ °C, correlates with inhomogeneous stresses in sol-gel films (owing to strong 25% shrinkage), as indicated indirectly from IR absorption. Based on the independent annealing behavior of the two PL signals and comparison with prior studies of sol-gel film properties, it appears that the Er is not directly involved in the Si emission. The spin-coating process presented herein is notable for enhancing light emission from bulk-type Si (estimated QE ~1%) in the absence of any patterning or p-n junctions.

## Acknowledgements

Partial support by the University of Jordan, the New Jersey Institute of Technology, and the U.S. National Renewable Energy Laboratory are gratefully acknowledged. To be presented at TMS 2012 Annual Meeting, Symposium on Recent Developments in Biological, Electronic, Functional and Structural Thin Films and Coatings (www.tms.org).